\documentclass[11pt]{article}
\usepackage{a4wide}
\usepackage{amstext,amsfonts,amsmath}

\begin{document}
\begin{flushright}
FTUV--04-0727  \quad IFIC--04--39\\
July 27, 2004
\end{flushright}

\begin{center}
\vspace{.7cm} {\Large {\bf Spacetime scale-invariant
 super-$p$-brane actions on enlarged superspaces and the geometry of
 $\kappa$-symmetry}}

\vspace{.7cm}

\bigskip

\renewcommand{\thefootnote}{\alph{footnote}}

\vskip 1cm

{\bf\large Jos\'e A. de Azc\'arraga}
${}^{\dag}$\footnote{E-mail: {\tt j.a.de.azcarraga@ific.uv.es}},
{\bf\large Jos\'e M. Izquierdo}
${}^{\ddag}$\footnote{E-mail: {\tt izquierd@fta.uva.es}},\\
{\bf\large C. Miquel-Espanya} ${}^{\dag}$
\footnote{E-mail: {\tt Cesar.Miquel@ific.uv.es}}

\vskip 0.4truecm

\vskip 0.2cm ${}^{\dag}$\ {\it Departamento de F\'{\i}sica
Te\'orica and IFIC (Centro Mixto CSIC-UVEG), 46100-Burjassot
(Valencia) Spain}

\vskip 0.2cm ${}^{\ddag}$\ {\it Departamento de F\'{\i}sica
Te\'orica,\\ Facultad de Ciencias, E-47011-Valladolid, Spain}

\vspace{3cm}

{\bf Abstract}

\end{center}

 We use the additional variables of suitably enlarged superspaces
 to write new actions for
 extended objects, with $\kappa$-symmetry, in such a way
 that the tension emerges from
 them as an integration constant. Our actions correspond to
 the spacetime scale-invariant ones previously considered
 by Bergshoeff {\it et al.} once the worldvolume forms introduced
 there are reinterpreted in terms of fields associated with the
 coordinates of the enlarged superspaces. It is shown that
 the $\kappa$-symmetry of the new actions is given by a certain type
 of right local transformations of the extended superspace
 groups. Further, we also show that the enlarged superspaces that
 allow for strictly invariant Wess-Zumino terms also lead to strict
 $\kappa$-invariance {\it i.e.}, the Lagrangian itself (not
 only the action) is both supersymmetry- and $\kappa$-invariant.

\begin{quotation}

\small

\end{quotation}

\newpage

\section{Introduction}
\setcounter{footnote}{0}

Super-$p$-brane actions in rigid superspace contain a kinetic
term, that supersymmetrizes the Nambu-Goto-Dirac action, plus a
Wess-Zumino (WZ) term. This last is the integral of a
supersymmetry quasi-invariant Lagrangian ({\it i.e.}, invariant
but for a total derivative) \cite{Ac87,Ev88}. The WZ term has to
be present, with the appropriate normalization factor, for the
complete  action to be invariant under  $\kappa$-symmetry (see
\cite{Ac87}
 for a general analysis). The quasi-invariance of the WZ terms
 leads to central terms in the algebra of charge densities and
 to topological `central' charges \cite{AGIT89}\footnote{The
 $p$-brane WZ terms are associated
 \cite{AzTo89} with $p+2$-cocycles on the standard graded translations
 algebra, and therefore produce extensions of the superalgebra of
 ($p$-dimensional) charge {\it densities} that have to be integrated
 (central extensions of the algebras of charges are directly produced by
 two-cocycles, the $p$=0 case).}(in connection with super-$p$-brane charges,
 see also \cite{SoTo-97,Hammer-97,ACIP-00,Hat-Sak-00,Ha-Pa-Smi-03}). These
 charges bypass Witten's `no-go' theorem \cite{Wi81} and allow for the partial
 breaking of rigid supersymmetry (PBRS) \cite{HuLiPo86}, as implied
 by the presence of $\kappa$-symmetry: WZ terms, $\kappa$-invariance
 of the action and PBRS are all related. In fact, it is possible
 to see all these phenomena already for the massive superparticle
 in \cite{Az-Lu-82} which, due to its WZ term, is the $p=0$ analogue
 \cite{AzLu83,Ev90} of the super-$p$-branes \cite{Ac87}; their
 tension $T$, of dimensions [$T$]=$ML^{-p}$ ($c$=1),
 corresponds to the superparticle mass $m$ for $p=0$, and one might
 rather write $T(p)$ for $T$ with $T(0)=m$. Furthermore, the appearance of
 the central mass term in the algebra of Noether charges \cite{Az-Lu-82}
 constitutes the simplest example of the {\it classical}
 `anomalies' that are present in the higher $p$ superbrane
 case, where $T$ enters \cite{Az-Iz-To-91}. Notice, nevertheless,
 that the properties of the $\kappa$-symmetry of the scalar branes
 \cite{Ac87} and of the massive superparticle \cite{Az-Lu-82,AzLu83} are
 different in so far as this last allows for a covariant fixing of
 $\kappa$-symmetry (as for other $D$-branes see
 \cite{Ag-Po-Schw-97,Ka-97,An-Gr-03}).

 The WZ term of the massive superparticle may be written in
 a strictly invariant form\footnote{In spite of being a
 somewhat self-contradicting terminology, we still refer to
 these invariant terms as WZ terms.} (see \cite{AICup}) by using
 a new coordinate that corresponds to the central generator of the,
 say, $D=4$, $N=2$ supersymmetry, the mass being the central charge.
 This is seen to be equivalent to the reverse of the dimensional
 reduction procedure that may be used to obtain the massive
 superparticle action from that of the massless superparticle,
 the mass appearing as the integration constant that fixes the
 value of the extra momentum component. For the scalar super-$p$-branes
 one may also obtain invariant WZ terms
 \cite{BeSe95,ACIP-00} by using suitable
  extensions\footnote{Throughout this paper the word `extension'
 is used in its (super)group/algebra extension theory meaning,
 {\it not} in a `N-extended' supersymmetry sense.}
 $\tilde \Sigma$ of the standard superspaces $\Sigma$ \cite{ACIP-00}.
 Moreover, for all $p\geq 0$ cases one may find that the
 contribution to the Noether charges resulting from the
 quasi-invariance of the original WZ terms, responsible
 for the topological charges \cite{AGIT89}, is provided
 by that coming from the additional extended superspace group
 variables once the WZ terms become invariant. Both contributions
 are equal, but have a different origin: one results from the
 quasi-invariance, the other from the additional variables
 that make the WZ terms invariant under the extended symmetry group.
 This reflects the connection between the quasi-invariance
 of the Lagrangians under symmetry groups and their (central)
 extensions (see also \cite{AICup} for the general theory).

 However, in spite of the above similarities with the massive
 superparticle, the derivation of the string or of a higher $p$
 superbrane action from a scale-invariant one, where $T$ does not enter,
 is more involved since $p\neq 0$. Indeed, for $p\geq 1$ it is not
 obvious how to derive by dimensional reduction
 the super-$p$-brane  action from that of a null $p$-brane in such a
 way that the tension appears as an integration constant.
 Two possible solutions to this problem were put forward
 in the past. The first \cite{AITb91} introduced a new variable $x_*(\tau)$
 depending on the first of the coordinates
  $\xi=(\tau,\sigma^i)$, $i=1,\dots, p$
  of the worldvolume $\cal W$, playing the role of the extra variable
 of the particle case. The second \cite{To92,BLT92} (see
 \cite{Be-To-98} for D-branes and also \cite{BeSoTo-98,DeGaHoSo-92})
 avoided using variables not locally
 defined on the worldvolume, but was based on the introduction of an
 independent $p$-form gauge potential $A(\xi)$,
 directly defined on $\cal W$, such that the variation of its
 field strength $dA(\xi)$ cancels exactly the variation of the WZ
 part of the Lagrangian. In both cases spacetime scale-invariant
 actions were written in such a way that the tension arose as
 an integration constant.

 In this paper we present a third possibility, although related to
 the second case above. We construct new rigid superspace
 actions for $p=1,2$ (although there is no obstacle
 for higher $p$) by adding worldvolume fields associated with
 the extra variables of suitably extended
 superspace groups, generically denoted  $\tilde \Sigma$.
 In this way, we adhere to the enlarged superspace variables/worldvolume
 fields correspondence principle for superbranes \cite {ACIP-00}, by
 which all branes worldvolume fields (and not only $x^\mu(\xi)$,
 $\theta^\alpha(\xi)$) originate from the (enlarged) superspace
 $\tilde\Sigma$ variables {\it i.e.}, are obtained from pull-backs
 by the map $\phi:\mathcal{W}\rightarrow \tilde\Sigma $ that
 locates the $p$-brane worldvolume in the enlarged
 superspace\footnote{The idea of a `fundamental symmetry between
 coordinates and fields' is explicitly stated in Berezin \cite{Be-79}
 and is implicit in earlier work of D. V. Volkov;
 see also \cite{Sch-et-al-80}.}. The rigid superspace
 $\tilde\Sigma$ is the group manifold of an extension of the standard
 supersymmetry group $\Sigma$ that is in general non-central (except in the
 case of the string Green algebra \cite{Gr89}) and that contains new odd,
 spinorial generators. These extended superspace group manifolds
 $\tilde\Sigma$ (see \cite{ACIP-00,BeSe95}) trivialize the
 Chevalley-Eilenberg (CE) cohomology $(p+2)-$cocycles on
 the ordinary supersymmetry algebra that characterize \cite{AzTo89} the
 quasi-invariant WZ terms of the scalar super-$p$-branes \cite{Ac87},
 making {\it Lie} (super)algebras \cite{BeSe95} of the original free
 differential algebras. The new
 actions also reduce to the ordinary super-$p$-brane actions
 when the field equations of the additional superspace variables,
 that appear only in the WZ term (see also \cite{ACIP-00} for
 other cases), are used. They may be seen to
 correspond to those in \cite{To92,BLT92} once the worldvolume
 $p$-form $A(\xi)$ introduced there is expressed as the pull-back
 of a $p$-form on the appropriate enlarged superspace, much in
 the same way that the world-volume Born-Infeld fields
 of the D-branes, directly introduced in their actions, may be obtained
 from forms on suitably enlarged superspaces \cite{ACIP-00,Saka}
 (for an analysis of the $D$-$p$-brane Noether charges see
 \cite{SoTo-97,Hammer-97,Be-To-98,ACIP-00}). The above
 fact is perhaps not surprising because for $p=1$ the
 actions of \cite{BLT92} can  be written in a Born-Infeld
 form, and the relation of $dA$ there with the invariant form
 $F$, $F=dA-b$, where $db=dF$ is also invariant (and a CE cocycle),
 is that of a Born-Infeld field \footnote{There are other examples where
 fields introduced initially `by hand' find an interpretation in terms of
 variables of a suitably enlarged superspace, as that of the three-form
 $A_3(x)$ field of $D=11$ supergravity
 \cite{CSJ-78,D'A+F-82,BAIPV-04}, that support the idea of an extended
 superspace variables/fields correspondence. That the
 three form $A_3$ can be `trivialized' is not surprising:
 the $D=11$ supermembrane WZ term is the pull-back to $\cal W$
 of the potential of the (CE four-cocycle on superspace) $dA_3$, and we know
 that the WZ terms may be trivialized on extended superspaces. The
 above \cite{D'A+F-82,BAIPV-04} trivializations are, however,
 different.}.

 It was early noticed that the fermionic gauge symmetry
 found in \cite{Az-Lu-82} for the massive
 and in \cite{Sie83} for the massless superparticle ($p=0$) cases,
 called $\kappa$-symmetry, had its origin in a kind of
 `right' local supersymmetry transformation (with one-half spinor
 independent parameters). This was exhibited
 \cite{Az-Lu-82,AzLu83,Az-Iz-To-91} by the fact that the fermionic constraints
 of the super-$p$-branes (half of which are first class, generating
 the local fermionic invariance that removes the unphysical fermionic
 degrees of freedom) correspond to  supercovariant derivatives
 (the "{\cal{$D$}'s") that generate the right supersymmetry
 transformations (see \cite{AlAz}), a fact also seen in
 \cite{Fry84} for the $p$=0 case and considered in a general
 framework in \cite{McA-00} for super-$p$-branes. Another
 treatment of $\kappa$-symmetry was given in
 a twistor-like doubly supersymmetry formulation of the massless
 superparticle action \cite{SoTkVo-89}, in which $\kappa$-symmetry
 was identified with proper time (worldline) supersymmetry. This was extended
 to super-$p$-branes (see, {\it e.g.} \cite{DeGaHoSo-92})
 in the superembedding approach (see \cite{BaSoToPaVo-95}
 and \cite{Sor-00} for a review and further references.)}.
 Here we exhibit the geometrical character of $\kappa$-symmetry
 as right local transformations when enlarged superspaces $\tilde\Sigma$
 are considered. We shall devote a special attention to the
 role and definition of $\kappa$-transformations for
 the new actions on these enlarged superspaces (explicitly
 for $p=0,1,2$). It will be shown that these actions have both strict
 supersymmetry and $\kappa$-symmetry, the latter being given
 in terms of the standard one-half spinor local transformations
 $\epsilon(\kappa)$ plus additional local bosonic
 $b(\kappa)$-translations.

The paper is organized as follows. In Sec. 2 we first review
 the massive particle case  and consider in Sec. 2.2 an action
 in the first order formalism (which is actually that of a massless
 superparticle in one more dimension) defined on the superspace of
 a centrally extended supersymmetry algebra, to recall how the mass
 appears. In Secs. 3 and 4 we consider the $p=1,2$ super-$p$-brane
 actions in enlarged superspaces $\tilde\Sigma$ on which the Wess-Zumino
 terms are strictly invariant under supersymmetry. We rewrite
 them as counterparts of the first order action of the
 superparticle, suitable for our purposes, and then present
 in Secs. 3.2 and 4.2 new actions for $p=1$ and $p=2$
 respectively in a way analogous to the $p=0$ case. Next we
 show in Secs. 3.3 and 4.3 that these actions are
 $\kappa$-invariant, and study the geometry of the $\kappa$-symmetry
 transformations, both fermionic and bosonic,
 in the context of the extended supersymmetry groups.

 We note that all considerations in this paper are made for actions
 on {\it rigid} superspaces which, in that case, are supergroup
 manifolds. For supergravity backgrounds (curved superspaces),
 $\kappa$-transformations may be related to one-half
 of the fermionic general coordinate transformations of standard
 superspace (pulled-back to the worldvolume) preserved by the
 brane action. The group-theoretical
 structure appears there when the supergravity constraints
 describing the standard rigid superspace are considered. We shall
 not discuss this here and refer instead to {\it e.g.}
 \cite{BaAzIzLu-02} and references therein, where the local
 symmetries of the dynamical supergravity-superbrane interacting
 system are also considered.

\section{The massive superparticle and its $\kappa$-symmetry}

\subsection{The massive superparticle action with invariant WZ term}

The action for the massive superparticle \cite{Az-Lu-82} is
\begin{equation}
I=\int d\tau\{-m\sqrt{-\omega_\mu\omega^\mu}+
 \lambda\epsilon_{IJ}C_{\alpha\beta}
 \theta^\alpha_I{\dot\theta}^\beta_J\} \label{A1}
\end{equation}
where $m$ is the mass of the particle, $I=1,2$ ($N$=2),
 $\epsilon_{IJ}=-\epsilon_{JI}$, $\epsilon_{12}=1$,
 $\theta_I$ are Majorana spinors, $\omega_\mu$ is given by
\begin{equation}
\omega_\mu={\dot x}_\mu+\theta^\alpha_I(C\Gamma_\mu)_{\alpha\beta}
 {\dot\theta}^\beta_I\label{A2}
\end{equation}
 and $\lambda$ is fixed so that the action has the required
 $\kappa$-invariance \cite{Az-Lu-82,AzLu83,Fry84}; we use mostly
 plus signature. The key feature of the WZ term in
 (\ref{A1}) is that it is the worldline expression of the
 superspace quasi-invariant one-form
 $\epsilon_{IJ}C_{\alpha\beta}\theta^\alpha_I d\theta^\beta_J$, the
 differential of which,
 $\epsilon_{IJ}C_{\alpha\beta} d\theta^\alpha_I\wedge
 d\theta^\beta_J$,
 is a {\it non-trivial} two-cocycle\footnote{A non-trivial
 ($p$+2)-cocycle is given by a closed and left-invariant
 ($p$+2)-form that it is not the exterior derivative of a
 {\it left-invariant} ($p$+1)-form.}
 (here, for the $N$=2 supersymmetry
 algebra cohomology; we shall omit wedge products henceforth). By
 construction, the potential WZ one-form
 $\epsilon_{IJ}C_{\alpha\beta}\theta^\alpha_Id\theta^\beta_J$
 cannot be an exact differential. This non-trivial two-cocycle
 property \cite{Az-Lu-82} is {\it the} property that determines
 all massive superparticle (0-brane) actions in any dimension $D$,
 much in the same way as the non-trivial higher order CE cocycles
 characterize \cite{AzTo89} (see also \cite{ACIP-00}) other $p>0$ branes.
 For instance, for $D=4$ the charge conjugation matrix $C$ is
 antisymmetric, and the bilinears  ${\bar \theta}_1\theta_2$,
 ${\bar \theta}_1\gamma^5\theta_2$, where $\theta_{1,2}$ are
 Majorana spinors, are symmetric under the exchange
 1$\leftrightarrow$2. This is already sufficient to discard
 $m{\bar\theta}{\dot \theta}$, $m{\bar\theta}\gamma^5{\dot \theta}$
 as $N$=1, $D$=4, $p$=0 WZ terms (they are total derivatives).
 Hence a $N$=2 superspace was used for the original massive
 superparticle action with WZ term \cite{Az-Lu-82} since it
 was specifically written for $D=4$.
 In contrast, since $C$ is symmetric for $D$=9,
 $m{\bar d\theta}d\theta$ is a non-trivial two-cocycle
 and $m{\bar \theta}{\dot \theta}$  a  suitable
 $N$=1, $D$=9 massive superparticle WZ term.
 Similarly, in the  $D$=10, IIA case, $\theta$
 is a Majorana spinor made out of two Majorana-Weyl ones of
 opposite chirality and ${\bar \theta}_1\gamma^{11}\theta_2$
 is antisymmetric; hence, $m{\bar d\theta}\gamma^{11}d\theta$
 is a non-trivial two-cocycle and $m{\bar \theta}\gamma^{11}{\dot\theta}$
 the $D$=10, IIA massive superparticle (actually, a D0-brane)
 WZ term. All other cases follow analogously.

 To look for the value of $\lambda$ in the action (\ref{A1})
 we shall first introduce a new coordinate $\varphi$
 in such a way that the second (WZ) term in (\ref{A1}) becomes
 invariant. The action (\ref{A1}) is built
 using the pull-backs of the  left-invariant Maurer-Cartan
 (MC) one-forms, $\Pi^\mu\equiv dx^\mu+\theta^\alpha_I
 (C\Gamma^\mu)_{\alpha\beta} d\theta^\beta_I$ and
 $\Pi^\alpha_I\equiv d\theta^\alpha_I$, defined on $N=2$ superspace
 $\Sigma^2$, to the particle worldline. These forms satisfy
 the MC equations of the standard $N=2$ supersymmetry algebra
\begin{equation}
d\Pi^\alpha_I=0\ ,\quad d\Pi^\mu=\Pi^\alpha_I
 (C\Gamma^\mu)_{\alpha\beta} \Pi^\beta_I\quad (I=1,2)\ . \label{A3}
\end{equation}
This superalgebra can be centrally extended, a fact that is equivalent
 to the existence of a closed invariant two-form on $\Sigma^2$,
 $\Omega_2=\epsilon_{IJ} C_{\alpha\beta}\Pi^\alpha_I\Pi^\beta_J$,
 which is not the exterior differential of a left-invariant one-form. However,
 its quasi-invariant potential one-form $\epsilon_{IJ}C_{\alpha\beta}
 \theta^\alpha_I\Pi^\beta_J$ can be supplemented with a piece
 $d\varphi$ in such a way that the transformation properties
 of $\varphi$ make the resulting one-form left-invariant. In this way, a
 larger superspace group (a central extension of the $N=2$
 superspace) is defined, on which the one-form
 $\Pi^\varphi\equiv d\varphi+\epsilon_{IJ}C_{\alpha\beta}\theta^\alpha_I
 \Pi^\beta_J$, $d\Pi^\varphi=\Omega_2$, is the MC form
 associated with the central charge coordinate $\varphi$ of the
 $N=2$ enlarged superspace $\tilde\Sigma^2$. The one-form
 $\Pi^\varphi$ can be pulled back to the worldline to define
 an invariant WZ term for the massive superparticle:
\begin{equation}
I=\int d\tau\{-m\sqrt{-\omega_\mu\omega^\mu}+
 \lambda\omega^\varphi\}\label{A4}
\end{equation}
where
\begin{equation}
\phi^*(\Pi^\varphi)\equiv \phi^*(d\varphi+\epsilon_{IJ}
 C_{\alpha\beta}\theta^\alpha_I\Pi^\beta_J)=({\dot\varphi}+
 \epsilon_{IJ} C_{\alpha\beta}\theta^\alpha_I{\dot\theta}^\beta_J)
 d\tau\equiv \omega^\varphi d\tau\ . \label{A5}
\end{equation}

 It is easy to see here that $\kappa$-symmetry is a right local
 supersymmetry translation depending of one-half of the parameters.
 The MC forms $\Pi^\alpha$, $\Pi^\mu$ of (\ref{A3}) and $\Pi^\varphi$
 of (\ref{A5}) are all left-invariant under rigid extended
 supersymmetry transformations. Under right
 local transformations, the MC forms transform as gauge fields.
 Indeed, the canonical left-invariant form  on a Lie group $G$,
 $g^{-1}dg=\omega^i\circ T_i$ (where $T_i$ is a basis of ${\cal G}$),
 transforms under local right
 translations $h$ of coordinates $h^i$ as
\begin{equation}
g^{-1}dg\longmapsto (gh)^{-1}d(gh)=h^{-1}(g^{-1}dg) h+
 h^{-1}dh \ , \label{A6}
\end{equation}
and the component MC forms $\omega^i$ as
\begin{equation}\label{new1}
\delta_{h}\omega^j\simeq dh^j + C^j_{ik}\omega^i h^k
\end{equation}
 So we do not need to write the explicit form of $\Pi^\alpha$,
 $\Pi^\mu$ and $\Pi^\varphi$ in terms of the $\tilde\Sigma^2$
 coordinates $\theta^\alpha$, $x^\mu$ and
 $\varphi$ to know the gauge
 transformation of the Lagrangian; it suffices to know the
 extended superalgebra structure constants.

Here we need the transformation of the MC one-forms under
 the local fermionic supertranslation
 $\epsilon^\alpha_I(\kappa)$,
\begin{equation}
\delta_\kappa \Pi^\alpha_I=d\epsilon^\alpha_I(\kappa)\ ,\quad
 \delta_\kappa\Pi^\mu=2\epsilon^\alpha_I(\kappa)
 (C\Gamma^\mu)_{\alpha\beta}\Pi^\beta_I\ ,\quad
 \delta_\kappa\Pi^\varphi=2\epsilon_{IJ}C_{\alpha\beta}
 \epsilon^\alpha_I(\kappa)\Pi^\beta_J\ .\label{A7}
\end{equation}
Using (\ref{A7}) in (\ref{A4}), one obtains
\begin{equation}
\delta I=\int d\tau 2\epsilon^\alpha_I(\kappa)\left\{
 m\frac{\omega_\mu}{\sqrt{-\omega^\nu\omega_\nu}}
 (C\Gamma^\mu)_{\alpha\beta}\delta_{IJ}+
 \lambda\epsilon_{IJ}C_{\alpha\beta}\right\}
 {\dot\theta}^\beta_J \ , \label{A8}
\end{equation}
It is then straightforward to check that if $\lambda=\pm m$ and
\begin{equation}
\epsilon^\alpha_I= \frac{\omega_\mu}
 {\sqrt{-\omega^\nu\omega_\nu}}{(\Gamma^\mu)^\alpha}_\gamma
 \kappa^\gamma_I\mp \epsilon_{IJ}\kappa^\alpha_J = \mp
 \epsilon_{IK}\left(\delta_{KJ}\delta^\alpha_\beta\pm\epsilon_{KJ}
 \frac{\omega_\mu}{\sqrt{-\omega^2}}
 \left(\Gamma^\mu\right)^\alpha_{\ \beta}\right)\kappa^\beta_J\ ,\label{A9}
\end{equation}
where the expression in brackets is twice a projector with one-half
 eigenvalues equal to zero (its second term squares to unity and is traceless),
 the variation of $I$ vanishes identically. Therefore the
 $\kappa$-invariant action is

\begin{equation}
I=\int d\tau\{-m\sqrt{-\omega^\mu\omega_\mu}\pm m\omega^\varphi \}
 \ .\label{A10}
\end{equation}
 It is interesting to note that the bosonic and fermionic first
 class constraints for the Lagrangian in eq. (\ref{A10}), that
 produce the Klein-Gordon and Dirac equation for the fields
 of the first quantized model \cite{Az-Lu-82,AzLu83,Fry84},
 give fields of equal mass (as required by supersymmetry)
 due to the presence of $\kappa$-symmetry, which
 requires $\lambda=\pm m$.

Let us now write the action in first order form by introducing
$p_\mu(\tau)$ and the einbein $E(\tau)=e(\tau)d\tau$:
\begin{equation}
I=\int d\tau\{ p^\mu\omega_\mu- \frac{1}{2}e(p^\mu
 p_\mu+m^2)\pm m\omega^\varphi\} \ .\label{A11}
\end{equation}
This action is classically equivalent to (\ref{A10}). Indeed, the
 $p^\mu$ equation gives  $p^\mu=\omega^\mu/e$.
 This can be substituted into the action, which now depends
 on $\omega^\mu,e,\omega^\varphi$. The $e$-dependence can be
 removed by using its Lagrange equation. Choosing the positive
 sign, $e=\frac{1}{m}\sqrt{-\omega_\mu\omega^\mu}$, and introducing
 this again into the action gives (\ref{A10}).

\subsection{The superparticle mass as an integration constant}

The new variable $\varphi$ appears in the action (\ref{A11})
 as a total  derivative, but if we write instead the action
\begin{equation}
I_*=\int d\tau \{p^\mu\omega_\mu-\frac{1}{2}e(p^\mu p_\mu+p_*^2) +
 p_*\omega^\varphi \}\label{A12}
\end{equation}
$\varphi$ is no longer dynamically trivial, and its Lagrange
 equation gives ${\dot p}_*=0$. Then, setting $p_*=\pm m$ and
 substituting in (\ref{A12}), one recovers (\ref{A11}): the
 mass of the particle can be viewed as an integration constant.
 By doing this, we restrict the variable $p_*$ to a
 particular {\it classical} solution, so that in moving to
 (\ref{A11}) we have lost a dynamical degree of freedom
 present in (\ref{A12}); obviously, the massive superparticle
 action (\ref{A11}) is not equivalent to the action
 (\ref{A12}). This is clear at the quantum
 level, where one has to integrate over all degrees of
 freedom. The same observation will apply to
 the higher order branes to be considered later.

The action $I_*$ of (\ref{A12}) can be re-interpreted
 as an action for a  massless superparticle in one dimension
 higher, if $\varphi$ is treated as the
 $(D+1)$-th new spatial coordinate and $p_*$ as its associated
 momentum. Inserting the $p^\mu$ and $p_*$ equations
\begin{equation}
 p^\mu=\frac{\omega^\mu}{e}\ ,\quad
 p_*=\frac{\omega^\varphi}{e}\quad ,\label{A13}
\end{equation}
into (\ref{A12}), we obtain
\begin{equation}
I_*=\int
d\tau\frac{1}{2e}\{\omega^\mu\omega_\mu+(\omega^\varphi)^2\}\equiv
\int d\tau \frac{1}{2e}\omega^{\tilde\mu}\omega_{\tilde\mu}\
.\label{A14}
\end{equation}

\section{The $N=1$, $p=1$ superstring case}

\subsection{Superstring actions in enlarged superspace}

As in the massive superparticle case, the Lagrangian of the
 $D=10$, $N=1$ superstring  has a quasi-invariant Wess-Zumino term,
 which also can be made strictly invariant  by enlarging the
 standard superspace to one $\tilde\Sigma$ with coordinates
 $x^\mu,\theta^\alpha,\varphi_\alpha$, the variables of the extended
 superspace group corresponding to the Green algebra \cite{Gr89}
 . Its MC equations are\footnote{In Sec. 3 all appropriate
 expressions have the chiral projector understood, so that all spinors
 are Majorana-Weyl ($\theta^\alpha\equiv \mathcal{P}_+\theta^\alpha$,
 $C\Gamma^\mu$ corresponds to $\sigma^\mu,$ etc).}
\begin{eqnarray}
d\Pi^\alpha &=& 0\ ,\nonumber\\
 d\Pi^\mu &=&\frac{1}{2}(C\Gamma^\mu)_{\alpha\beta}
 \Pi^\alpha\Pi^\beta\ ,\nonumber\\
 d\Pi_\alpha &=& (C\Gamma_\mu)_{\alpha\beta}
 \Pi^\mu\Pi^\beta \ . \label{B1}
\end{eqnarray}
Notice that $\Pi^\alpha$ and $\Pi_\alpha$ are associated with the
 different (Majorana-Weyl) fermionic variables $\theta^\alpha$ and
 $\varphi_\alpha$ and that are hence unrelated (no charge conjugation
 matrix is used to move the index $\alpha$). Since we shall
 use them to construct the new action, we give below
 the MC equations (\cite{ACIP-00}, eq. (48)) of a
 generalization of the above Green algebra
\begin{eqnarray}
d\Pi^\alpha &=& 0\ ,\quad d\Pi^\mu =\frac{1}{2}(C\Gamma^\mu)_{\alpha\beta}
 \Pi^\alpha\Pi^\beta\ ,\nonumber\\
 d\Pi^\mu_\varphi &=& \frac{1}{2}(C\Gamma^\mu)_{\alpha\beta}
 \Pi^\alpha\Pi^\beta\ ,\nonumber\\
 d\Pi_\alpha &=& (C\Gamma_\mu)_{\alpha\beta}\Pi^\mu\Pi^\beta+
 (C\Gamma_\mu)_{\alpha\beta}\Pi^\mu_\varphi \Pi^\beta \ .
 \label{B11a}
\end{eqnarray}
Adding the Lorentz automorphisms part, eqs. (\ref{B11a}) determine
 an enlarged $N=1$ superPoincar\'{e} group $\widetilde{sP}$ parametrized by
 $(x^\mu,\theta^\alpha,\varphi_\alpha,\varphi_\mu,
 \alpha^{\mu\nu})$. Thus,
 $(x^\mu,\theta^\alpha,\varphi_\alpha,\varphi_\mu)$ parametrize
 an {\it extended $D=10$ superspace} group
 manifold $\tilde\Sigma\ (=\widetilde{sP}/L)$.

 Using the MC forms in (\ref{B11a}), we may construct a
 strictly invariant WZ term for the superstring following
 \cite{Sie94}. Writing
 $\phi^*(\Pi^\mu)=\Pi^\mu_i d\xi^i$, $\phi^*(\Pi^\mu_\varphi)
 =\Pi^\mu_{\varphi i} d\xi^i$, $\phi^*(\Pi^\alpha)=\Pi^\alpha_i d\xi^i$,
 $\phi^*(\Pi_\alpha)=\Pi_{\alpha i}
 d\xi^i$ for the pull-backs of the MC forms on the worldsheet,
 and $M_{ij}=\Pi^\mu_i\Pi_{\mu j}$ for the induced metric on $\mathcal{W}$,
 the invariant superstring action is (see \cite{ACIP-00})
\begin{equation}
 I=-\int d^2\xi \{ T\sqrt{-det M}+\lambda\epsilon^{ij}
 (\Pi_{\mu i}\Pi^\mu_{\varphi j}-\frac{1}{2}
 \Pi_{\alpha i}\Pi^\alpha_j)\}\ ,\label{B2}
\end{equation}
 where again $\lambda$ is to be fixed by $\kappa$-symmetry. Note that in
 this action the new variables $\varphi_\alpha$,
 $\varphi^\mu$ ({\it i.e.}, those beyond the standard superspace $\Sigma$
 coordinates $(x^\mu,\theta^\alpha)$) appear only in the
 invariant WZ term and through a total derivative. This is evident once
 one computes its exterior differential with the help of (\ref{B1}),
 since the result involves only $\Pi^\alpha$ and $\Pi^\mu$:
\begin{equation}
d(\Pi_\mu\Pi^\mu_\varphi-\frac{1}{2}\Pi_\alpha\Pi^\alpha)\equiv
dB=-(C\Gamma_\mu)_{\alpha\beta}
 \Pi^\mu\Pi^\beta\Pi^\alpha\ .\label{B3}
\end{equation}

To show that $\kappa$-symmetry is a right local supersymmetry
depending on one-half of the fermionic parameters,
 we need to know the form of these
 transformations. Either taking
\begin{eqnarray}
\delta_\kappa \theta^\alpha & = &
 \epsilon^\alpha (\kappa) \nonumber \\
 \delta_\kappa x^\mu & = &  -\frac{1}{2}
 (C\Gamma^\mu)_{\alpha\beta} \theta^\alpha
 \epsilon^\beta(\kappa)  \nonumber\\
 \delta_\kappa \varphi^\mu & = & -\frac{1}{2}
 (C\Gamma^\mu)_{\alpha\beta} \theta^\alpha
 \epsilon^\beta(\kappa) \nonumber\\
 \delta_\kappa \varphi_\alpha &=& -\frac{1}{2}
 (C\Gamma^\mu)_{\alpha\beta} x_\mu\epsilon^\beta(\kappa)
 -\frac{1}{2} (C\Gamma^\mu)_{\alpha\beta} \varphi_\mu
 \epsilon^\beta(\kappa)\nonumber\\
 & &+\frac{1}{6} (C\Gamma_\mu)_{\alpha\beta}(C\Gamma^\mu)_{\gamma\delta}
 \theta^{(\gamma}\theta^{\beta)}
 \epsilon^\delta(\kappa)
\end{eqnarray}
(see \cite{ACIP-00} for the extended superspace group law
corresponding to the superalgebra (\ref{B11a})) or directly from
the transformation properties of the left invariant MC one-forms
under local right supertranslations, eq. (\ref{new1}), we obtain
\begin{eqnarray}
\delta_\kappa \Pi^\alpha &=& d(\epsilon^\alpha
 (\kappa))\ ,\nonumber\\
 \delta_\kappa \Pi^\mu &=& -(C\Gamma^\mu)_{\alpha\beta}
 \Pi^\alpha\epsilon^\beta (\kappa)\ ,\nonumber\\
 \delta_\kappa \Pi^\mu_\varphi &=& - (C\Gamma^\mu)_{\alpha\beta}
 \Pi^\alpha\epsilon^\beta (\kappa)\nonumber\\
 \delta_\kappa \Pi_\alpha &=& -(C\Gamma_\mu)_{\alpha\beta} \Pi^\mu
 \epsilon^\beta (\kappa) - (C\Gamma_\mu)_{\alpha\beta}
 \Pi^\mu_\varphi \epsilon^\beta (\kappa)\ . \label{B4}
\end{eqnarray}
The $\kappa$-variation of the action
 may now be calculated from eq. (\ref{B4}) as follows. For the kinetic
 (`Dirac') part one may use
 $\delta \det M=\det M M^{ij}\delta M_{ij}$. That of the WZ term is
 calculated by noticing that the second term in (\ref{B2}) is the
 pull-back $\phi^*(\lambda B)$ to the worldsheet, where $B$ is
 defined in (\ref{B3}), and that the variations (\ref{B4}) are given by Lie
 derivatives $L_X$ where the vector field has only the
 $\epsilon^\alpha(\kappa)$ components,
 $i_X(\Pi^\alpha)=\epsilon^\alpha(\kappa)$
 (the other contractions are zero). Then, the relevant term of
 $\delta_{\kappa} B=L_X B=(i_X d+d i_X)B$ in the action
 is $i_X dB$ which, by (\ref{B3}), is given by
 $2(C\Gamma_\mu)_{\alpha\beta}\Pi^\mu\epsilon^\beta(\kappa)\Pi^\alpha$.
 It is useful to introduce the worldvolume Dirac matrices and their
 antisymmetrized products by the definitions
\begin{equation}
\Gamma_i\equiv \Pi^\mu_i\Gamma_\mu \quad
(\{\Gamma_i,\Gamma_j\}=2M_{ij}) \quad , \quad
\Gamma_{ij}=\Gamma_{\mu\nu}\Pi^\mu_i\Pi^\nu_j\quad .
\end{equation}
Then,
\begin{equation}
\delta_\kappa I=-\int d^2\xi\{ -T\sqrt{-det M}M^{ij}
 (C\Gamma_i)_{\alpha\beta}\Pi^\alpha_j\epsilon^\beta(\kappa)
 +2\lambda\epsilon^{ij} (C\Gamma_i)_{\alpha\beta}
 \epsilon^\beta(\kappa)\Pi^\alpha_j\}\quad .\label{B5}
\end{equation}
If we now set
\begin{equation}
\epsilon^\alpha(\kappa)=\frac{1}{2}\left(\delta^\alpha_\beta
 \pm\frac{1}{2} \frac{\epsilon^{ij}{(\Gamma_{ij})^\alpha}_\beta}
 {\sqrt{-det M}}\right) \kappa^\beta \ ,\label{B6}
\end{equation}
 the matrix acting on $\kappa$ is again a projection
 operator since $\frac{1}{2}\frac{\epsilon^{ij}
 {(\Gamma_{ij})^\alpha}_\beta} {\sqrt{-det M}}$ squares to unity,
 and, being traceless, $\epsilon(\kappa)$ depends on
 half of the $\kappa$ parameters. Then, choosing
 $\lambda=\pm\frac{1}{2}T$ we see that $\delta_\kappa
 I=0$, and the $\kappa$-invariant action reads
\begin{equation}
I=-\int d^2\xi \left\{ T\sqrt{-det M} \pm\frac{T}{2}
 \epsilon^{ij}\left(\Pi_{\mu i}\Pi^\mu_{\varphi j}-\frac{1}{2}
 \Pi_{\alpha i}\Pi^\alpha_j\right)\right\}
\ .\label{B7}
\end{equation}

We can now write the action (\ref{B7}) in a `first order-like'
form analogous to (\ref{A11}) by introducing
$p^{\mu\nu}(\xi)=p^{[\mu\nu]}(\xi)$ and $e(\xi)$, which
nevertheless lacks the interpretation in
 terms of the canonical formalism that is applicable to the particle
 case:
\begin{equation}
 I=-\int d^2\xi\frac{1}{2}\left\{ p^{\mu\nu}\epsilon^{ij}\Pi_{\mu
 i} \Pi_{\nu j}+\frac{1}{2}e(p^{\mu\nu}p_{\mu\nu}+2T^2)
 \pm T\epsilon^{ij}\left(\Pi_{\mu i}\Pi^\mu_{\varphi j}-\frac{1}{2}
 \Pi_{\alpha i}\Pi^\alpha_j\right)\right\}\, . \label{B8}
\end{equation}
 In differential form language, and using $B$
 (eq. (\ref{B3})), $I$ reads
\begin{equation}
I=-\int\frac{1}{2} \{p^{\mu\nu}\Pi_{\mu}\Pi_{\nu}+
 \frac{1}{2}Ep^{\mu\nu}p_{\mu\nu}+T^2 E \pm
 TB\} \quad ,\label{B8f}
\end{equation}
where

\begin{equation}
e(\xi)=\frac{1}{2}\epsilon^{ij}E_{ij}(\xi)\quad ,
\quad E(\xi)=\frac{1}{2}E_{ij}(\xi) d\xi ^i
d\xi^j\quad .\label{ebrane}
\end{equation}

It is easy to prove that this action is classically equivalent
 to (\ref{B7}).  First, the Euler-Lagrange (E-L) equation
 for $p^{\mu\nu}$ tells us
 that $p_{\mu\nu}=-\frac{1}{e} \epsilon^{ij}\Pi_{\mu i}
 \Pi_{\nu j}$. Substituting this into the action (\ref{B8}),
 solving for $e$, $e=\frac{1}{T}\sqrt{-det M}$, and substituting
 again yields (\ref{B7}). In the process, one uses the identity
\begin{equation}
\epsilon^{ij}\epsilon^{rl}\Pi_{\mu i}\Pi_{\nu j}
 \Pi^\mu_r\Pi^\nu_l = 2 det M \ .\label{B9}
\end{equation}

  We conclude this subsection with a remark on one more
 form of the action (\ref{B8f}). The evident (`rest-like system')
 solution
 $p_{(a)(b)} =  T \delta^{++}_{[(a)}\delta_{(b)]}^{--}\;$
 ($\delta^{\pm\pm}_{(a)}= \delta^0_{(a)}\pm\delta^{(D-1)}_{(a)})\;$
 of the constraint $p_{\mu\nu} p^{\mu\nu}=-2T^2$ (provided by
 the E-L equation for $e$) may be made covariant by means of the
 Lorentz group matrix
\begin{equation}
\Lambda_\mu{}^{(a)} = [\; 1/2 (u_\mu^{++} +
u_\mu^{--})\;,\; u_\mu^i \; ,\; 1/2 (u_\mu^{++} - u_\mu^{--}) \; ]
\quad \in\quad SO(1,D-1) \; , \label{Lambda-har}
\end{equation}
where $u_\mu^{++}$ ($u_\mu^{--}$) corresponds to $u^0+u^{(D-1)}$
($u^0-u^{(D-1)}$) and $i=1,\dots,D-2$. Inserting the solution
\begin{equation}
p_{\mu\nu} = p_{(a)(b)} \Lambda_\mu^{(a)} \Lambda_\nu^{(b)} =  T
u^{++}_{[\mu } u_{\nu ]}^{--}\; , \qquad \label{p-solution}
\end{equation}
into the action (\ref{B8f}) we arrive at
\begin{equation}
I= -T \int (\frac{1}{2}\Pi^\mu u_\mu^{++} \; \Pi^\nu u_\nu^{--}
\pm B) \; . \qquad \label{harm-action}
\end{equation}
This is the so-called Lorentz-harmonics formulation of the
superstring action \cite{BaZhe-92} (see also
\cite{BaSoVo-95,BaSoToPaVo-95}), but now with a strictly invariant
WZ term.

\subsection{A new action in the superstring case}

In Sec. 2.2 we substituted a new momentum $p_*$ for $m$ to obtain the
 generalized action (\ref{A12}). Here we shall do something similar,
 by adding  to the action a new worldsheet field (also denoted
 $p_*$ but obviously different, $p_*=p_*(\xi)$) that will replace the tension.
 This new, spacetime scale-invariant form $I_*$ of the action (\ref{B8f}) has
 the expression
\begin{equation}
I_*=-\int d^2\xi\{ \frac{1}{2}p^{\mu\nu}\epsilon^{ij}\Pi_{\mu i}
 \Pi_{\nu j}+\frac{1}{4}e(p^{\mu\nu}p_{\mu\nu}+8p_* ^2) +p_* \Phi\}
 \ ,\label{B10}
\end{equation}
where, for convenience, we have written $\phi^*(B)=\frac{1}{2}
 B_{ij}d\xi^i d\xi^j=\Phi d^2\xi$
\begin{equation}
\Phi\equiv\frac{1}{2}\epsilon^{ij}B_{ij}=\epsilon^{ij}(\Pi_{\mu i}\Pi^\mu_{\varphi j}-
 \frac{1}{2} \Pi_{\alpha i}\Pi^\alpha_j)
 \label{B101}
\end{equation}
(recall that $\Pi_\alpha$ and $\Pi^\alpha$ are unrelated).

Let us show now that $T$ in (\ref{B8}) appears as an integration
 constant (see the comment after (\ref{A12})). To this aim,
 let us first compute the E-L
 equations for the enlarged superspace worldsheet fields.
 These result from setting equal to zero the coefficients
 of $i_X\Pi^\alpha$, $i_X\Pi^\mu$,
 $i_X\Pi_\alpha$, $i_X \Pi^\mu_\varphi$ in a general variation
 $\delta=L_X=di_X +i_Xd\sim i_X d$ of the
 Lagrangian, where $i_X$ is the inner derivation with respect to the
 vector field that determines an arbitrary variation $\delta$
\begin{equation}
X=\Pi(\delta)^\alpha \,D_\alpha + \Pi(\delta)^\mu\,D_\mu
 +\Pi(\delta)_\alpha \,D^\alpha +
 \Pi(\delta)^\mu_\varphi\,D^\varphi_\mu \quad ,
 \label{ELcov}
\end{equation}
where the $D$'s are the left-invariant vector fields on $\tilde
\Sigma$ (so that $\Pi^A(D_B)=\delta^A_B$) and the $\Pi(\delta)$'s
are the corresponding MC forms in which $d\theta, dx^\mu,
d\varphi_\alpha, d\varphi^\mu$ have been replaced by the
variations $\delta\theta, \delta x^\mu, \delta\varphi_\alpha,
\delta\varphi^\mu$. Adding the equations for $p^{\mu\nu}$, $p_*$
and $E$, the complete set of E-L equations in differential form is
given by
\begin{eqnarray}
& & -\frac{1}{2} d(p_*\Pi_\alpha)+ p^{\mu\nu} \Pi_\nu
 (C\Gamma_\mu)_{\alpha\beta}\Pi^\beta -\frac{1}{2}p_*(C\Gamma_\mu)_{\alpha\beta}
 \Pi^\beta \Pi^\mu_\varphi  \nonumber\\
 & & \qquad\qquad\qquad\qquad\qquad\qquad- \frac{3}{2}p_*
 (C\Gamma_\mu)_{\alpha\beta} \Pi^\mu \Pi^\beta =0\label{el7}\\
 & &  -d(p^{\mu\nu}\Pi_\nu)-d(p_*\Pi^\mu_\varphi)-\frac{1}{2}p_*
 (C\Gamma^\mu)_{\alpha\beta} \Pi^\alpha\Pi^\beta = 0 \label{el4}\\
 & & dp_* \Pi^\alpha = 0 \label{el6} \\
 & & dp_* \Pi^\mu= 0 \label{el5} \\
& & \Pi^\mu\Pi^\nu+E p^{\mu\nu}= 0 \label{el1}\\
 & &\Pi_\mu\Pi^\mu_\varphi-\frac{1}{2}\Pi_\alpha\Pi^\alpha
 +4Ep_*=0 \label{el2}\\
 & &  p^{\mu\nu}p_{\mu\nu}+8p_*^2=0 \label{el3}\ .
\end{eqnarray}

Equation (\ref{el5}) in worldsheet coordinates is
 $\epsilon^{ij}\partial_i p \Pi^\mu_j=0$, which contracted
 with $\Pi_{\mu k}$ gives
\begin{equation}
\epsilon^{ij}\partial_i p M_{jk} =0\ .
\end{equation}
If the induced metric $M_{ij}$ is non-degenerate, one immediately
 concludes that $p_*$ is a constant;  if not, eqs.
 (\ref{el3}) and (\ref{el1}) (or (\ref{pmunu}) below) tell us
 that $p_*=0$ necessarily,  so in any case $p_*$ is a constant that
 we can set equal to $\pm \frac{T}{2}$ (with $T=0$ for $p_*=0$).
 If we use this solution in $I_*$, the result
 is (\ref{B8}). When $p^{\mu\nu}$ and $p_*$ are eliminated using
 their E-L equations, the resulting action is
\begin{equation}
I_*=\int d^2\xi\frac{1}{e}\left\{ \frac{det M}{2}+
 \frac{\Phi^2}{8} \right\}\quad,\label{B11}
\end{equation}
where $\Phi$ is given in (\ref{B101}). This recovers the action of
\cite{To92} although there $\Phi$ involves the field strength of
an independent worldsheet field $A_i(\xi)$. In the present
framework, $A_i(\xi)$ is expressed as
 a composite of fields associated with the enlarged superspace
 coordinates in the spirit of the mentioned extended superspace
 group variables/fields correspondence principle
 for branes \cite{ACIP-00}.

To see that (\ref{B10}) leads to (\ref{B11}) explicitly,
 let us solve the algebraic equations for $p^{\mu\nu}$ and
 $p_*$,
\begin{eqnarray}
p_{\mu\nu}&=&-\frac{1}{e}\epsilon^{ij}\Pi_{\mu i}\Pi_{\nu j}
\label{pmunu}\\
  p_*&=&-\frac{1}{4e}\epsilon^{ij}\left(\Pi_{\mu i}
 \Pi^\mu_{\varphi j}-\frac{1}{2}\Pi_{\alpha i}\Pi^\alpha_j\right)
 \ . \label{p}
\end{eqnarray}
Inserting eqs. (\ref{pmunu}), (\ref{p}) into $I_*$ (eq.
(\ref{B10})) one obtains
\begin{eqnarray}
I_*&=&-\int d^{p+1}\xi \Bigg{\{}-\frac{1}{2e}\epsilon^{ij}
 \Pi^\mu_i\Pi^\nu_j \epsilon^{kl}\Pi_{\mu
 k}\Pi_{\nu l} +\frac{1}{4e}\epsilon^{ij}\Pi^\mu_i\Pi^\nu_j
 \epsilon^{kl}\Pi_{\mu k}\Pi_{\nu l}\nonumber\\
 &\ &-\frac{1}{4e}\epsilon^{ij}\left(\Pi_{\mu i}
 \Pi^\mu_{\varphi j}-\frac{1}{2}\Pi_{\alpha i}
 \Pi^\alpha_j \right)\epsilon^{kl}\left( \Pi_{\mu k}
 \Pi^\mu_{\varphi l}-\frac{1}{2}\Pi_{\alpha k}
 \Pi^\alpha_l\right)\nonumber\\
 &\ &+\frac{1}{8e}\epsilon^{ij}\left( \Pi_{\mu i}
 \Pi^\mu_{\varphi j}-\frac{1}{2}\Pi_{\alpha i}
 \Pi^\alpha_j \right)\epsilon^{kl}\left( \Pi_{\mu k}
 \Pi^\mu_{\varphi l}-\frac{1}{2}\Pi_{\alpha k}
 \Pi^\alpha_l\right)\Bigg{\}}\nonumber\\
 &=& \int d^{p+1}\xi\left\{\frac{1}{4e}\epsilon^{ij}
 \epsilon^{kl}M_{ik}M_{jl}+\frac{1}{8e}\Phi^2\right\}\ ,
\end{eqnarray}
which is eq. (\ref{B11}). We note, for later use, the E-L for e,

\begin{equation}\label{e1-E-L}
  \det M +\Phi^2/4=0\quad .
\end{equation}

\subsection{$\kappa$-invariance of the new actions}

We now exhibit the $\kappa$-symmetry of the new superstring
 action on the enlarged superspace
 both in the forms of eq. (\ref{B10}) and (\ref{B11}).

\bigskip

\noindent{\it (a) $\kappa$-invariance of the action (\ref{B10})}

\bigskip

To see that this action is invariant under the $\kappa$-symmetry
 transformations, one may prove that the equations for
 $\theta^\alpha$, eqs. (\ref{el7}), are not independent from
 the other E-L equations. This, by virtue of the second Noether's
 theorem, reflects the presence of a fermionic gauge symmetry
 of the action.

The first step is to realize that
\begin{equation}
p^{\mu\nu}\Gamma_{\mu\nu}+ 4p_*\label{kproy}
\end{equation}
is $(8p_*)$ times a projector. One has to compute the square of
 the matrix $p^{\mu\nu}\Gamma_{\mu\nu}$, which has a piece
 proportional to $p_{\mu\nu}p^{\mu\nu}$ which by eq.
 (\ref{el3}) is equal to $-8p_*^2$, but also has contributions of the
 form $\Gamma^{\mu\nu\rho\sigma} p_{\mu\nu}p_{\rho\sigma}$ and
 $\Gamma^{\mu\nu}p_{\mu\rho}{p^\rho}_\nu$. These two types of
 contributions vanish, due to index symmetry arguments, if one takes
 into account eq. (\ref{el1}) in the form
\begin{equation}
p^{\mu\nu} =-\frac{1}{e} \epsilon^{ij}
 \Pi^\mu_i \Pi^\nu_j\ .\label{el1s}
\end{equation}
Next, after using $dp_*=0$ and the last equation of
 (\ref{B11a}), the fermionic equation (\ref{el7}) reads
\begin{equation}
\left[2p_*(C\Gamma_\mu)_{\alpha\beta}\Pi^\mu+
 p^{\mu\nu}(C\Gamma_\nu)_{\alpha\beta} \Pi_\mu\right]
 \Pi^\beta=0\ .
\end{equation}
To show that some of these equations (actually half of them) are
 trivially satisfied we multiply them by (\ref{kproy}).
 Again, there are terms that vanish due to index symmetry considerations once
 (\ref{el1s}) is used. The only two surviving terms are
 proportional to $p_*^2\Gamma_\mu\Pi^\mu$ and ${p^\rho}_\nu
 p^{\mu\nu}\Gamma_\rho \Pi_\mu$ respectively. The latter may,
 for instance, be re-expressed as a term of the form
 $p_{\mu\nu}p^{\mu\nu}\Gamma_\rho\Pi^\rho$ by making the
 substitution (\ref{el1s}), then using that the
 antisymmetrization of three worldsheet indices
 vanishes, and finally going back to an expression
 involving $p_{\mu\nu}$. Then both terms cancel
 each other due to eq. (\ref{el3}). This shows the presence
 of the standard number of fermionic gauge ($\kappa$-)symmetries
 under which the action is invariant.

\bigskip

\noindent{\it (b) $\kappa$-invariance of the action (\ref{B11})}

\bigskip

 We now find explicitly the form of the gauge
 $\kappa$-transformations under which the action (\ref{B11})
 is invariant, in order to exhibit its geometrical nature
 as right local supersymmetries
 in the context of our enlarged superspace. We could do
 this in the case of action (\ref{B10}), but the
 transformation of the auxiliary field $p^{\mu\nu}(\xi)$ is
 non-trivial, and this obscures the
 geometrical interpretation.

 Under a local right translation of parameters
 $\epsilon^{\alpha}(\kappa)$, $b^\mu(\kappa)$ associated with the
 right transformations of $\theta^\alpha$, $\varphi^\mu$, the
 variation of the MC forms is read from the structure
 constants of the extended algebra,
\begin{eqnarray}
\delta_\kappa\Pi^\alpha&=& d\epsilon^\alpha(\kappa)\label{C2}\\
 \delta_\kappa\Pi^\mu&=&-(C\Gamma^\mu)_{\alpha\beta}
 \Pi^\alpha\epsilon^\beta(\kappa)\\
 \delta_\kappa\Pi^\mu_\varphi&=&-(C\Gamma^\mu)_{\alpha\beta}
 \Pi^\alpha\epsilon^\beta(\kappa) + db^\mu(\kappa)\\
 \delta_\kappa\Pi_\alpha&=&-(C\Gamma_\mu)_{\alpha\beta}\Pi^\mu
 \epsilon^\beta(\kappa)-(C\Gamma_\mu)_{\alpha\beta}
 \Pi^\mu_\varphi\epsilon^\beta(\kappa)+(C\Gamma^\mu)_{\alpha\beta}
 \Pi^\beta b_\mu(\kappa)\, .\label{C3}
\end{eqnarray}
 Using these variations we may now compute
 $\delta M_{jl}=-(C\Gamma_l)_{\alpha\beta}{\Pi^\alpha}_j\epsilon^\beta(\kappa)+
 (l\leftrightarrow j)$, from which we obtain
 \begin{eqnarray}
 \delta_\kappa \det M & = & \epsilon^{ij}\epsilon^{kl}
 M_{ik}\delta_\kappa M_{jl}=2\epsilon^{ij}\epsilon^{kl}M_{ik}\Pi_{\mu j}
 \delta_\kappa\Pi^\mu_l \nonumber\\
 & = & -2\epsilon^{ij}\epsilon^{kl}M_{ik}
 (C\Gamma_j)_{\alpha\beta}\Pi^\alpha_l\epsilon^\beta(\kappa)\quad .
 \label{var-det-1}
\end{eqnarray}
 There are no terms in $\delta_{\kappa}\det M$ containing $b^\mu(\kappa)$,
 exactly as in the standard superstring action. But since $\Phi$
 does depend on the the new variables, there are contributions both from
 $\epsilon^\alpha(\kappa)$ and $b^\mu(\kappa)$ to $\delta_\kappa\Phi$.
 The $\epsilon^\alpha(\kappa)$ contribution gives
\begin{eqnarray}
\delta_{\epsilon(\kappa)} \Phi& = & \epsilon^{ij}\Big{(}
 -(C\Gamma_\mu)_{\alpha\beta}\Pi^\alpha_j\epsilon^\beta(\kappa)
 \Pi^\mu_{\varphi j}-\Pi_{\mu i}(C\Gamma^\mu)_{\alpha\beta}
 \Pi^\alpha_j\epsilon^\beta(\kappa)\nonumber\\
 &\ &+\frac{1}{2}(C\Gamma_\mu)_{\alpha\beta}\Pi^\mu_i
 \epsilon^\beta(\kappa)\Pi^\alpha_j+\frac{1}{2}
 (C\Gamma_\mu)_{\alpha\beta}\Pi^\mu_{\varphi i}
 \epsilon^\beta(\kappa)\Pi^\alpha_j -\frac{1}{2}\Pi_{\alpha i}\partial_j
 \epsilon^\alpha(\kappa)\Big{)}\nonumber\\
 & = & \epsilon^{ij}[2(C\Gamma_\mu)_{\alpha\beta}
 \Pi^\mu_i\epsilon^\beta(\kappa)\Pi^\alpha_j-\frac{1}{2}\partial_j
 (\Pi_{\alpha i}\epsilon^\alpha(\kappa))] \quad . \label{C1}
\end{eqnarray}
 In the standard superstring action the total derivative term
 in eq. (\ref{C1}) is dealt with integrating by parts, which is not
 possible here. To cancel it we shall use instead the $b^\mu(\kappa)$
 part of the transformation in (\ref{C2}) - (\ref{C3})
 associated with $\varphi^\mu$, which is
\begin{eqnarray}
\delta_{b(\kappa)}\Phi & = & \epsilon^{ij}\left(\Pi_{\mu i}
 \partial_jb^\mu(\kappa)-\frac{1}{2}(C\Gamma^\mu)_{\alpha\beta}
 b_\mu(\kappa)\Pi_i^\beta\Pi_j^\alpha\right)\nonumber\\
 & = & \epsilon^{ij}\partial_j(\Pi_{\mu i}b^\mu(\kappa))
\end{eqnarray}

Let us assume that there is a $b^\mu(\kappa)$ such that
 $\Pi_{\mu i}b^\mu(\kappa) = \frac{1}{2} \Pi_{\alpha
i}\epsilon^\alpha(\kappa)$.
 Then,
\begin{equation}
 \delta_\kappa\Phi  =2\epsilon^{ij}(C\Gamma_i)_{\alpha\beta}
 \epsilon^\beta(\kappa)\Pi^\alpha_j\quad .
 \label{var-Phi}
\end{equation}
 Using eqs. (\ref{var-det-1}) and (\ref{var-Phi}) we obtain
\begin{equation}
\delta_\kappa\left(\det M +\frac{1}{4}\Phi^2\right)
 =-2\epsilon^{ij}\epsilon^{kl} M_{ik}(C\Gamma_j)_{\alpha\beta}
 \Pi^\alpha_l\epsilon^\beta(\kappa) +\Phi\epsilon^{ij}
 (C\Gamma_i)_{\alpha\beta} \epsilon^\beta(\kappa)
 \Pi^\alpha_j\quad , \label{CN}
\end{equation}
 and we see that if \footnote{For $\Phi\neq 0$, eq.
 (\ref{C5}) gives $\epsilon^\beta(\kappa)=\frac{\Phi}{e}
 \left(\delta^\beta_\gamma-\epsilon^{ij}
 \frac{(\Gamma_{ij})^\beta_{\ \gamma}}{\Phi}\right)
 \kappa^\gamma$ which exhibits its projector structure:
 the second term squares to unity on account of
 $(\epsilon^{ij}\Gamma_{ij})^2 = -(2!)^2\det M$ and eq. (\ref{e1-E-L}).
 Since $(\Gamma_{ij})^\beta_{\ \gamma}$ is traceless, half
 of the parameters are removed.}
\begin{equation}
\epsilon^\beta(\kappa)=\frac{1}{e} \left(\Phi\delta^\beta_\gamma -
 \epsilon^{ij}(\Gamma_{ij})^{\ \beta}_{\gamma}\right)
 \kappa^\gamma\label{C5} \ ,
\end{equation}
then
\begin{eqnarray}
\delta_\kappa\left(\det M +\frac{1}{4}\Phi^2\right) & = &
 -\frac{2}{e}\epsilon^{ij}\epsilon^{kl}M_{ik}
 (C\Gamma_j)_{\alpha\beta}\Pi^\alpha_l\Phi\,\kappa^\beta\nonumber\\
 & &+\frac{2}{e}\epsilon^{ij}\epsilon^{kl}M_{ik}
 (C\Gamma_j)_{\alpha\beta}\Pi^\alpha_l  \epsilon^{rs}
 (\Gamma_{rs})_{\ \gamma}^{\beta}\kappa^\gamma\nonumber\\
 & & +\frac{\Phi^2}{e}\epsilon^{ij}(C\Gamma_i)_{\alpha\beta}
 \kappa^\beta\Pi^\alpha_j-\frac{1}{e}\Phi\epsilon^{ij}(C\Gamma_i)_{\alpha\beta}
 \epsilon^{kl}(\Gamma_{kl})_{\ \gamma}^{\beta}
 \kappa^\gamma\Pi^\alpha_j\nonumber\\
 & = &-\frac{4}{e}\epsilon^{ij}(C\Gamma_i)_{\alpha\beta}
 \Pi^\alpha_j\kappa^\beta\left(\det M + \frac{1}{4}\Phi^2\right)\quad ,
\end{eqnarray}
where the terms linear in $\Phi$ cancel each other. Therefore, the
variation of the Lagrangian density is
\begin{eqnarray}
\delta_\kappa\left\{-\frac{1}{2e}\left(\det M
 +\frac{1}{4}\Phi^2\right)\right\} & = &  \nonumber \\
 \frac{1}{2e^2}(\delta_\kappa e)\left(\det M +\frac{1}{4}\Phi^2\right)&
 +&\frac{2}{e^2}\epsilon^{ij}(C\Gamma_i)_{\alpha\beta}\Pi^\alpha_j
 \kappa^\beta\left(\det M +\frac{1}{4}\Phi^2\right) \, ,
\end{eqnarray}
which is equal to zero if
\begin{equation}
\delta_{\kappa} e=-4\epsilon^{ij}(C\Gamma_i)_{\alpha\beta}
 \Pi^\alpha_j\kappa^\beta \quad . \label{63}
\end{equation}

Let us return to the question of solving
\begin{equation}
\Pi_{\mu i}b^\mu(\kappa) = \frac{1}{2}\Pi_{\alpha
 i}\epsilon^\alpha(\kappa) = \frac{\Pi_{\alpha i}}
 {2e}\left(\Phi\kappa^\alpha - \epsilon^{jk}\Gamma_{jk\ \
 \beta}^{\quad\alpha}\kappa^\beta\right) \ .\label{64}
\end{equation}
One might take
\begin{equation}
b^\mu(\kappa)=\frac{1}{2}\Pi^\mu_i M^{ij}\Pi_{\alpha
 j}\epsilon^\alpha(\kappa) \quad . \label{65}
\end{equation}
because then
\begin{equation}
\Pi_{\mu s}b^{\mu}(\kappa)=\frac{1}{2}M_{si}M^{ij}\Pi_{\alpha j}
 \epsilon^\alpha(\kappa) = \frac{1}{2}\delta^j_s \Pi_{\alpha j}
 \epsilon^\alpha(\kappa) = \frac{1}{2}\Pi_{\alpha s}
 \epsilon^\alpha(\kappa)\ ,
\end{equation}
 which holds provided $M^{ij}$ exists. This is not always the
 case because some solutions of the E-L equations (those corresponding
 to the tensionless string case) have degenerate induced
 metric. But these solutions also imply $\Phi=0$ by virtue of the
 E-L equation for $e$ (eq. (\ref{e1-E-L})) so in that case we can write
 (\ref{64}) with $\Phi=0$. The solution of the resulting
 equation is
\begin{equation}
b^\mu (\kappa)= \frac{1}{e}
 (\Gamma^\mu_j)^\alpha_{\ \beta}
 \Pi_{l\alpha} \kappa^\beta\epsilon^{jl}\quad . \label{69}
\end{equation}
 This completes the proof of the $\kappa$-symmetry for the
 form (\ref{B11}) of the action. We notice that the projector
 nature of the bracket in (\ref{C5}) (as it was the case
 for (\ref{A9}) and will be for (\ref{brane-project}) below)
 requires using the E-L eqs. for $e$, eq. (\ref{ebrane}), although the
 $\kappa$-symmetry of the action itself does not. The full set
 of $\kappa$-symmetry
 transformations are expressed by eqs. (\ref{C2})-(\ref{C3})
 and (\ref{63}) with the fermionic $\epsilon(\kappa)$ and the
 bosonic $b^\mu(\kappa)$
 given by (\ref{C5}) and (\ref{65}) or (\ref{69}) respectively,
 and again we see that they correspond to right local
 transformations depending on a fermionic parameter
 $\kappa$\footnote{The presence of additional bosonic gauge
 transformations beyond reparametrization invariance
 is a feature of superspaces with additional bosonic coordinates;
 {\it cf.} \cite{Az-Lu-82} and \cite{BaLu-99, BaAzPiVa-03}
 for the case of the $\Sigma^{(n(n-1)/2|n)}$ superspaces, and
 references therein. The additional
 bosonic gauge `$b$'-symmetries there should not be confused, however, with the
 $\kappa$-symmetry bosonic transformations discussed in the present paper,
 since the former depend on bosonic parameters that are
 $\kappa$-independent; they are, rather, bosonic superpartners of
 $\kappa$-symmetry.}.
 Furthermore, we see that the $\kappa$-symmetry of (\ref{B11})
 is strict or manifest: the Lagrangian, and not only
 the action, is $\kappa$-invariant.

\section{The $N=1$ $p=2$ $D=11$ supermembrane case}

\subsection{The supermembrane with invariant WZ term}

As we did in the last section, we consider here the rigid supermembrane
 action with an invariant WZ term written as a product of MC
 one-forms on an enlarged superspace. The particular superspace
 needed to do this is the group manifold (parametrized by $\theta^\alpha$,
 $x^\mu$, $\varphi^{\mu\nu}$, $\varphi^{\mu\alpha}$, $\varphi_{\alpha\beta}$)
 \cite{ACIP-00} associated with the extended supersymmetry
 algebra defined by the MC equations
\begin{eqnarray}
d\Pi^\alpha &=& 0\quad , \quad d \Pi^\mu = \frac{1}{2} (C\Gamma^\mu)_{\alpha\beta}
 \Pi^\alpha \Pi^\beta \quad ,\nonumber\\
 d \Pi^{\mu\nu} &=& \frac{1}{2}
 (C\Gamma^{\mu\nu})_{\alpha\beta}  \Pi^\alpha \Pi^\beta\quad ,\nonumber\\
 d\Pi_{\mu\alpha} &=& (C\Gamma_{\nu\mu})_{\alpha\beta}
 \Pi^\nu\Pi^\beta + (C\Gamma^\nu)_{\alpha\beta} \Pi_{\nu\mu}
 \Pi^\beta\quad ,\nonumber\\
 d\Pi_{\alpha\beta} &=&  -\frac{1}{2}
 (C\Gamma_{\mu\nu})_{\alpha\beta} \Pi^\mu \Pi^\nu
 -\frac{1}{2} (C\Gamma^\mu)_{\alpha\beta} \Pi_{\mu\nu}
 \Pi^\nu \nonumber\\
 & & + \frac{1}{4} (C\Gamma^\mu)_{\alpha\beta}
 \Pi_{\mu\delta} \Pi^\delta +(C\Gamma^\mu)_{\delta\alpha}
 \Pi_{\mu\beta}\Pi^\delta +(C\Gamma^\mu)_{\delta\beta}
 \Pi_{\mu\alpha} \Pi^\delta\ .
 \label{Me1}
\end{eqnarray}
Using these equations, it is straightforward to see that the
 CE three-cocycle on the standard
 supersymmetry algebra that characterizes
 the membrane, $(C\Gamma_{\mu\nu})_{\alpha\beta}
 \Pi^\mu \Pi^\nu \Pi^\alpha \Pi^\beta$, can be written as
 $dB$, where $B$ is now the three-form \cite{BeSe95,ACIP-00}
\begin{equation}
B=\frac{2}{3}\Pi^\mu\Pi^\nu \Pi_{\mu\nu} + \frac{3}{5}
 \Pi^\mu \Pi^\alpha \Pi_{\mu\alpha}- \frac{2}{15}
 \Pi_{\alpha\beta} \Pi^\alpha\Pi^\beta\ . \label{Me2}
\end{equation}
Since the three-form $B$ is constructed out of MC forms
 of a (enlarged) supersymmetry algebra \cite{BeSe95} it
 provides an invariant WZ term.

 Let us now show that the $\kappa$-symmetry gauge transformations
 are also here a special type of right local transformations with
 parameters $\epsilon^\alpha(\kappa)$. From the change (see \cite{ACIP-00})
 of the group variables $\theta^\alpha$, $x^\mu$, $\varphi^{\mu\nu}$,
 $\varphi_{\mu \alpha}$ and $\varphi_{\alpha\beta}$ under
 local, right transformations $\delta_\kappa$, we obtain
\begin{eqnarray}
\delta_\kappa \theta^\alpha &=& \epsilon^\alpha(\kappa)\quad ,\nonumber\\
 \delta_\kappa x^\mu &=& -\frac{1}{2} (C\Gamma^\mu)_{\alpha\beta}
 \theta^\alpha \epsilon^\beta(\kappa)\quad ,\nonumber \\
 \delta_\kappa \varphi_{\mu\nu} &=& -\frac{1}{2}
 (C\Gamma_{\mu\nu})_{\alpha\beta} \theta^\alpha
 \epsilon^\beta(\kappa)\quad ,\nonumber \\
 \delta_\kappa \varphi_{\mu\alpha} &=& -\frac{1}{2}
 (C\Gamma^\nu)_{\alpha\beta} \varphi_{\mu\nu}
 \epsilon^\beta(\kappa)-\frac{1}{2}
 (C\Gamma_{\mu\nu})_{\alpha\beta} x^\nu
 \epsilon^\beta(\kappa) \nonumber\\
 & & +\frac{1}{12} (C\Gamma^\nu)_{\alpha\beta}
 (C\Gamma_{\mu\nu})_{\gamma\delta} \theta^\beta \theta^\gamma
 \epsilon^\delta(\kappa) +\frac{1}{12} (C\Gamma^\nu)_{\gamma\delta}
 (C\Gamma_{\mu\nu})_{\alpha\beta} \theta^\beta \theta^\gamma
 \epsilon^\delta(\kappa) \quad ,\nonumber
 \end{eqnarray}
\begin{eqnarray}
 \delta_\kappa \varphi_{\alpha\beta} &=&
 -(C\Gamma^\mu)_{\beta\gamma} \varphi_{\mu\alpha}
 \epsilon^\gamma(\kappa) -\frac{1}{8} (C\Gamma^\mu)_{\alpha\beta}
 \varphi_{\mu\gamma} \epsilon^\gamma(\kappa) \nonumber\\
 & & -\frac{1}{48} (C\Gamma^\mu)_{\alpha\beta}
 (C\Gamma^\nu)_{\gamma\delta} \varphi_{\mu\nu} \theta^\gamma
 \epsilon^\delta(\kappa) +\frac{1}{48} (C\Gamma^\mu)_{\alpha\beta}
 (C\Gamma_{\mu\nu})_{\gamma\delta} x^\nu \theta^\gamma
 \epsilon^\delta(\kappa) \nonumber\\
 & & -\frac{1}{6} (C\Gamma^\mu)_{\alpha\gamma}
 (C\Gamma^\nu)_{\beta\delta} \varphi_{\mu\nu} \theta^\gamma
 \epsilon^\delta(\kappa) -\frac{1}{6} (C\Gamma^\mu)_{\alpha\gamma}
 (C\Gamma_{\mu\nu})_{\beta\delta} x^\nu \theta^\gamma
 \epsilon^\delta(\kappa) \nonumber\\
 & &+ \frac{1}{12} (C\Gamma^\mu)_{\gamma\delta}
 (C\Gamma_{\mu\nu})_{\alpha\beta} x^\nu \theta^\gamma
 \epsilon^\delta(\kappa) \ .\label{Me3}
\end{eqnarray}
From eq. (\ref{Me3}) or from the usual transformation properties of
 the left-invariant MC one-forms
 under right transformations (eqs. (\ref{Me1}), (\ref{new1})), it follows that
\begin{eqnarray}
\delta_\kappa \Pi^\alpha &=& d\epsilon^\alpha(\kappa)\quad, \quad \delta_\kappa \Pi^\mu
 =-(C\Gamma^\mu)_{\alpha\beta}
 \Pi^\alpha \epsilon^\beta(\kappa)\quad , \nonumber\\
 \delta_\kappa \Pi^{\mu\nu} &=&-(C\Gamma^{\mu\nu})_{\alpha\beta}
 \Pi^\alpha \epsilon^\beta(\kappa)\quad ,\nonumber\\
 \delta_\kappa \Pi_{\mu\alpha} &=&-(C\Gamma_{\nu\mu})_{\alpha\beta}
 \Pi^\nu \epsilon^\beta (\kappa) -(C\Gamma^\nu)_{\alpha\beta}
 \Pi_{\nu\mu}\epsilon^\beta (\kappa)\quad ,\nonumber\\
 \delta_\kappa \Pi_{\alpha\beta} &=& -\frac{1}{4}
 (C\Gamma^\mu)_{\alpha\beta} \Pi_{\mu\delta}
 \epsilon^\delta(\kappa) - (C\Gamma^\mu)_{\delta\alpha}
 \Pi_{\mu\beta} \epsilon^\delta(\kappa)\nonumber\\
 & & -(C\Gamma^\mu)_{\delta\beta} \Pi_{\mu\alpha}
 \epsilon^\delta(\kappa) \quad ,\label{Me4}
\end{eqnarray}
for an appropriate $\epsilon^\alpha(\kappa)$ to be
 determined from the invariance of the action.

 Let us start by computing $\delta_\kappa B$. To do that, one may
 use (\ref{Me2}) and (\ref{Me4}) or notice that the right
 transformations are generated by the left-invariant vector
 fields (dual to the MC forms in (\ref{Me1})). Thus, $\delta_\kappa$
 is nothing but the Lie derivative $L_X$ with respect to a vector field
 $X=X(\kappa)$ such that $i_X\Pi=0$ for all $\Pi$'s in eq. (\ref{Me1}) but for
 $i_X\Pi^\alpha\equiv\Pi^\alpha(X)=\epsilon^\alpha(\kappa)$, so that only
 $\epsilon^\alpha(\kappa)$ enters in the components of X. Then,
\begin{eqnarray}
\delta_\kappa B &=& L_X B = i_X dB + d i_XB\nonumber\\
 &=& 2(C\Gamma_{\mu\nu})_{\alpha\beta}\Pi^\mu\Pi^\nu
 \epsilon^\alpha(\kappa) \Pi^\beta \nonumber\\
 & & + d\left( -\frac{3}{5}\Pi^\mu\epsilon^\alpha(\kappa)
 \Pi_{\mu\alpha} +\frac{4}{15}\Pi_{\alpha\beta}
 \epsilon^\alpha(\kappa) \Pi^\beta \right). \label{Me5}
\end{eqnarray}
Let the strictly invariant supermembrane action be given by ({\it
cf.} eq. (\ref{B2}) for $p$=1)
\begin{equation}
 I=-\int d^3\xi \left\{ T\sqrt{-det M}+\lambda\frac{1}{3!}\epsilon^{ijk}
 B_{ijk}\right\}\ , \label{Me5a}
\end{equation}
 where $M_{ij}= \Pi^\mu_i\Pi_{\mu j}$, $i,j=0,1,2$, $B_{ijk}$ are the
  coordinates of $\phi^*(B)=B(\xi)= \frac{1}{3!}B_{ijk}d\xi^i d\xi^j d\xi^k$,
  and again $\lambda$ is a constant to be fixed. Using
 the second equation in (\ref{Me4}) and eq. (\ref{Me5})
 (ignoring the total derivative term) we obtain
\begin{eqnarray}
\delta_\kappa I &=& -\int d^3\xi \left\{  -T \sqrt{-det M}
 (C\Gamma^l)_{\alpha\beta} \Pi^\alpha_l
 \left[\epsilon^\beta(\kappa) \right.\right. \nonumber\\
 & & + \left.\left.\frac{2\lambda}{3T \sqrt{-det M}} \epsilon^{ijk}
 {(\Gamma_{ijk})^\beta}_\gamma
 \epsilon^\gamma(\kappa)\right]\right\} \quad .\label{Me6}
\end{eqnarray}
 The expression between square brackets in (\ref{Me6}) can be written
 as ${P^\beta}_\gamma \epsilon^\gamma(\kappa)$, where
 $\frac{1}{2} {P^\beta}_\gamma$ is a projector that projects
 into half the spinor space if the square of the (traceless) matrix
\begin{equation}
\frac{2\lambda}{3T\sqrt{-det M}}\epsilon^{ijk} \Gamma_{ijk}
 \label{Me7}
\end{equation}
is the unit matrix. This is the case when $\lambda=\pm
 \frac{T}{4}$. Then one constructs
 $\epsilon(\kappa)$ as for $p=1$, and gets
\begin{equation}
\epsilon^\alpha(\kappa)=\frac{1}{2}\left(\delta^\alpha_\beta
 \mp\frac{1}{6\sqrt{-\det M}}\epsilon^{ijk}
 ({\Gamma_{ijk})^\alpha}_\beta\right)\kappa^\beta \quad .
\end{equation}
We remark that we differ from \cite{BeSe95} in that the new
 coordinates of the appropriate enlarged superspace group
 (\cite{ACIP-00}, eqs. (63)-(67)) are not inert under
 $\kappa$-symmetry, as seen in eqs. (\ref{Me3}) or (\ref{Me4}).
 We also note that we could have cancelled the second term in
 (\ref{Me5}) to obtain strict
 $\kappa$-invariance by adding to the transformations in
 (\ref{Me4}) a term in $b^{\mu\nu}(\kappa)$. We will do
 this in Sec. 4.3, where this bosonic contribution to
$\kappa$-symmetry will turn out to be necessary.

In order to obtain below the new action from that in eq.
(\ref{Me5a}), we give its `first order-like' formulation ({\it
cf.} eq. (\ref{B8}) for $p=1$) by introducing $p_{\mu\nu\rho}$ and
$e$ :
\begin{equation}
I= -\int d^3\xi \left\{ \frac{1}{3!} \epsilon^{ijk} p_{\mu\nu\rho}
 \Pi^\mu_i \Pi^\nu_j \Pi^\rho_k +e(p^{\mu\nu\rho} p_{\mu\nu\rho}
 +6 T^2)\pm \frac{T}{4} \Phi \right\}\ , \label{Me8}
\end{equation}
where ({\it cf.} eq. (\ref{B101}))
\begin{equation}\label{Bijk}
\Phi=(1/3!)\,\epsilon^{ijk}B_{ijk}\quad .
\end{equation}
This action is classically equivalent to (\ref{Me5a}) with
 $\lambda=\pm \frac{T}{4}$, as can be checked by eliminating
 $p_{\mu\nu\rho}$ and $e$ via their algebraic E-L equations.

\subsection{A new $p=2$ action in enlarged superspace}

Equation (\ref{Me8}) suggests introducing a new action by
 the following expression
\begin{equation}
I_* = -\int d^3\xi \left\{ \frac{1}{3!} \epsilon^{ijk}
 p_{\mu\nu\rho} \Pi^\mu_i \Pi^\nu_j \Pi^\rho_k +
 e(p^{\mu\nu\rho} p_{\mu\nu\rho} +96 p_*^2)+
 p_* \Phi \right\}\ . \label{Me8a}
\end{equation}
The E-L equations are computed as in Sec 3.2, eq. (\ref{ELcov})
and below. To show that $dp_*=0$ so that $p_*$ is constant it is
sufficient to use the E-L equations corresponding to
$\varphi^{\mu\nu}$. Specifically, we only need in the variation
$\delta I_*$ the coefficient accompanying $i_X\Pi^{\mu\nu}$.
 Since $\Pi^{\mu\nu}$ appears only in $B$ (or $\Phi$, eqs. (\ref{Me2}),
 (\ref{Bijk})), we only require
\begin{equation}
L_X(p_*B) = i_X d(p_*B) + d(p_*i_XB)\simeq i_X d(p_*B) \quad ,
\label{Me10}
\end{equation}
where the total derivative is ignored under the integral sign.
From the explicit form of $B$, we find that the $i_X\Pi_{\mu\nu}$
relevant term is
\begin{equation}
-\frac{2}{3} dp_*\Pi^\mu\Pi^\nu (i_X \Pi_{\mu\nu}) \label{Me11}
\end{equation}
so that, moving from differential forms to worldvolume fields,
 the $\varphi^{\mu\nu}$-associated E-L equation reads
\begin{equation}
\epsilon^{ijk}\partial_i p_* \Pi^\mu_j \Pi^\nu_k =0\ . \label{Me12}
\end{equation}
If we now contract this expression with $\Pi_{\mu r}
 \Pi_{\nu s}$, we obtain
\begin{equation}
\epsilon^{ijk}\partial_i p_* M_{jr}M_{ks} =0\ . \label{Me13}
\end{equation}
Then, as in the case of the string, if $det M=0$ the $\delta e$
 and $\delta p^{\mu\nu\rho}$ equations
\begin{equation}
p_{\mu\nu\rho}p^{\mu\nu\rho}+96p_*^2=0 \ , \quad
p_{\mu\nu\rho}=-\frac{1}{12e}\epsilon^{ijk}\Pi_{\mu i}\Pi_{\nu
j}\Pi_{\rho k}\ ,
\end{equation}
together with the fact that
\begin{equation}
  \det M= \frac{1}{3!}\epsilon^{ijk}\epsilon^{rst}M_{ir}M_{js}M_{kt} ,
  \label{det-M2}
\end{equation}
  tell us that $p_*=0$;
 if $det M\neq 0$ we may contract (\ref{Me13}) with $M^{rr{'}}
 M^{ss{'}}$, to arrive at $dp_*=0$. Thus, $p_*$ is a
 constant in any case. Setting $p_*=\pm \frac{T}{4}$ and
 introducing it into the action (\ref{Me8a}), one recovers the
 standard one, (\ref{Me8}).

We now write the action that results from using the algebraic
 equations for the $\delta p_*$ and $\delta p^{\mu\nu\rho}$
 variations in eq. (\ref{Me8a}). It is given by
\begin{equation}
I_*= \int d^3\xi \frac{1}{24 e} \left\{ det M+
 \frac{\Phi^2}{16}\right\}\ .\label{Me14}
\end{equation}
to be compared with that for the
 superstring, eq. (\ref{B11}). The E-L eq. for $e$ is

\begin{equation}\label{e2-E-L}
  \det M +\Phi^2/16=0  \quad .
\end{equation}

\subsection{$\kappa$-symmetry of the new action}

The $\kappa$-symmetry for the new form (\ref{Me8a}) of the action
 goes along the lines described in Sec. 3.3(a). We are more
 interested here, however, in the proof for the action
 (\ref{Me14}) (the case of Sec. 3.3(b) above), since this allows us to
 give more easily the explicit variations of the enlarged superspace
 variables and to check that these variations are again
 right local transformations depending on $\epsilon(\kappa)$.
 The novelty here is that, for $p=2$, they also depend on
 $b^{\mu\nu}(\kappa)$, $b^{\mu\nu}$ being the infinitesimal
 parameter associated with the new variable $\varphi^{\mu\nu}$.
 So we now add to (\ref{Me4}) the transformations that
 include the new $b^{\mu\nu}$ parameter; the $\kappa$-symmetry
 vector field now has non-zero components
 $\delta_\kappa\theta^\alpha = \epsilon^\alpha(\kappa)$,
 $\delta_\kappa\varphi^{\mu\nu}=b^{\mu\nu}(\kappa)$. This gives
\begin{eqnarray}
\delta_\kappa \Pi^\alpha &=& d\epsilon^\alpha(\kappa)\quad ,\nonumber\\
\delta_\kappa \Pi^\mu& =&-(C\Gamma^\mu)_{\alpha\beta}
 \Pi^\alpha \epsilon^\beta(\kappa) \quad ,\nonumber\\
 \delta_\kappa \Pi^{\mu\nu} &=& d b^{\mu\nu}(\kappa)
 -(C\Gamma^{\mu\nu})_{\alpha\beta} \Pi^\alpha \epsilon^\beta
 (\kappa)\quad ,\nonumber\\
 \delta_\kappa \Pi_{\mu\alpha} &=&(C\Gamma^\nu)_{\alpha\beta}
 b_{\nu\mu}(\kappa) \Pi^\beta
 -(C\Gamma_{\nu\mu})_{\alpha\beta}
 \Pi^\nu \epsilon^\beta (\kappa)
 -(C\Gamma^\nu)_{\alpha\beta}\Pi_{\nu\mu}
 \epsilon^\beta (\kappa)\quad ,\nonumber\\
 \delta_\kappa \Pi_{\alpha\beta} &=& - \frac{1}{2}
 (C\Gamma^\mu)_{\alpha\beta} b_{\mu\nu}(\kappa) \Pi^\nu
 -\frac{1}{4} (C\Gamma^\mu)_{\alpha\beta} \Pi_{\mu\delta}
 \epsilon^\delta(\kappa)\nonumber\\
 & & - (C\Gamma^\mu)_{\delta\alpha} \Pi_{\mu\beta}
 \epsilon^\delta(\kappa) -(C\Gamma^\mu)_{\delta\beta}
 \Pi_{\mu\alpha} \epsilon^\delta(\kappa)\ . \label{Me15}
\end{eqnarray}
Note that the variations of $\Pi^\alpha$ and $\Pi^\mu$ are
 exactly the same as in the standard, unextended algebra case.
 This implies, in particular, that the variation of $det\, M$
 in eq. (\ref{det-M2}) is given by ({\it cf.} (\ref{var-det-1}))
\begin{equation}
\delta_\kappa det\, M = -\epsilon^{ijk} \epsilon^{rst}
 M_{ir} M_{js} (C\Gamma_k)_{\alpha\beta}\Pi^\alpha_t
 \epsilon^\beta(\kappa)\quad ,
 \label{Me16}
\end{equation}
 since $\delta \det M= \frac{1}{2!}\epsilon^{ijk} \epsilon^{rst}
 M_{ir} M_{js} \delta M_{kt}$. The variation of $\Phi$
 is now given by ({\it cf.} (\ref{Me5}))
\begin{eqnarray}
\delta_\kappa\Phi &=& \epsilon^{ijk}\partial_i \left(
 \frac{3}{2} \Pi^\mu_j \Pi^\nu_k b_{\mu\nu}(\kappa)-
 \frac{3}{5} \Pi^\mu_j \epsilon^\alpha(\kappa)
 \Pi_{\mu \alpha \kappa}
 +\frac{4}{15} \Pi_{\alpha\beta j}
 \epsilon^\alpha(\kappa) \Pi^\beta_k \right) \nonumber\\
 & & + 2\epsilon^{ijk}
 (C\Gamma_{ij})_{\alpha\beta}\epsilon^\alpha(\kappa)
 \Pi^\beta_k \quad .
 \label{Me17}
\end{eqnarray}
However, in contrast with the action in Sec. 4.1, the total
derivative in (\ref{Me17}) cannot be ignored because it does not
produce a total derivative in the action (\ref{Me14}), since it
depends on $\Phi^2$ and $e$. This is precisely why a non-vanishing
$b^{\mu\nu}(\kappa)$ is now needed: its role is to cancel the
first term in (\ref{Me17}). At the same time, we see that the use
of enlarged superspaces not only produces strictly invariant WZ
terms; it also leads to strict $\kappa$-invariance.

If we assume that we have found a $b^{\mu\nu}(\kappa)$ such
 that the variation of $\Phi$ in eq. (\ref{Me17}) reduces to
 its second term, one checks that writing
\begin{equation}
\epsilon^\alpha(\kappa) = \frac{1}{e} \left(\Phi
 \delta^\alpha_\beta + \frac{2}{3} \epsilon^{ijk}
 {(\Gamma_{ijk})^\alpha}_\beta\right) \kappa^\beta\label{Me18}
\end{equation}
the variation of the terms between brackets in
 (\ref{Me14}) is given by
\begin{equation}
\delta_\kappa \left( det M+\frac{\Phi^2}{16} \right) = \frac{4}{e}
\epsilon^{ijk}(C\Gamma_{ij})_{\alpha\beta}
 \kappa^\alpha \Pi^\beta_k \left( det M+\frac{\Phi^2}{16}
 \right)\ ,\label{Me19}
\end{equation}
so that choosing suitably the variation of the auxiliary field
$e$,
\begin{equation}
\delta_\kappa e = 4 \epsilon^{ijk}(C\Gamma_{ij})_{\alpha\beta}
 \kappa^\alpha \Pi^\beta_k\ ,\label{Me20}
\end{equation}
the action remains invariant.

Again, for $\Phi\neq 0$ eq. (\ref{Me18}) has a projector structure,

\begin{equation}\label{brane-project}
\epsilon^\alpha(\kappa)=\frac{\Phi}{e}\left(\delta^\alpha_\beta+
  \frac{2}{3\Phi}\epsilon^{ijk}(\Gamma_{ijk})^\alpha_{\ \beta}\right)
  \kappa^\beta
\end{equation}
which follows from $(\epsilon^{ijk}\Gamma_{ijk})^2=-(3!)^2 \det M$
 and eq. (\ref{e2-E-L}).

To complete the proof of $\kappa$-invariance, we now have to
 show that the equation
\begin{equation}
\epsilon^{ijk}\left( \frac{3}{2} \Pi^\mu_j \Pi^\nu_k
 b_{\mu\nu}(\kappa) -\frac{3}{5} \Pi^\mu_j
 \epsilon^\alpha(\kappa) \Pi_{\mu \alpha k} +\frac{4}{15}
 \Pi_{\alpha\beta j}\epsilon^\alpha(\kappa) \Pi^\beta_k \right)=0
 \label{Me21}
\end{equation}
 does indeed have a solution for $b^{\mu\nu}$ when
 $\epsilon^\alpha(\kappa)$ is given by eq. (\ref{Me18}). As in
 Sec. 3.3, we shall consider separately the
 $\det M\neq 0$ and $\det M=0$ cases. When $M_{ij}$ is non-degenerate
 eq. (\ref{Me21}) has the solution
\begin{equation}
 b^{\mu\nu}(\kappa) = \frac{1}{3 detM} \Pi^\mu_r \Pi^\nu_s
 \epsilon^{rst} M_{tl} \epsilon^{lmn}
 \left(\frac{3}{5} \Pi^\rho_m \epsilon^\alpha(\kappa)
 \Pi_{\rho \alpha n} -\frac{4}{15} \Pi_{\alpha\beta m}
 \epsilon^\alpha(\kappa) \Pi^\beta_n \right)\ . \label{Me22}
\end{equation}
 We also have to worry about the $\kappa$-variation of
 configurations with $det M=0$, because these appear in the space of
 solutions of the E-L equations of the action (\ref{Me14}), when
 eq. (\ref{e2-E-L}) gives $\Phi=0$. In this case we have to
 solve eq. (\ref{Me21}) for an $\epsilon^\alpha(\kappa)$ given by
 eq. (\ref{Me18}) for $\Phi=0$. It is then seen that
\begin{equation}
b_{\mu\nu}(\kappa)= \frac{2}{3e}\epsilon^{ijk}\left(
 -\frac{6}{5}\Pi^\rho_j\Pi_{\rho\alpha k}+\frac{8}{15}
 \Pi_{\alpha\gamma j} \Pi^\gamma_k \right)({\Gamma_{\mu\nu
 i})^\alpha}_\beta \kappa^\beta\ , \label{Me23}
\end{equation}
 where $\Gamma_{\mu\nu k} = \Gamma_{\mu\nu \rho}\Pi^\rho_k$,
 cancels the first term in eq. (\ref{Me17}), and the rest of the
 proof follows as for $\Phi\neq 0$.
 So the action (\ref{Me14}) is $\kappa$-invariant, the variation
 of the Lagrangian components being given by
 (\ref{Me15}), (\ref{Me18}), (\ref{Me20}) and
 (\ref{Me22}) or (\ref{Me23}).

\section{Conclusions}

We have obtained new rigid superspace actions explicitly for
$p=1,2$ super-$p$-branes,
 starting from the usual superstring and supermembrane ones in which
 their WZ terms have been rewritten in a strictly invariant form
 by using MC forms defined on suitably extended superspace
 groups \cite{BeSe95,ACIP-00}. The procedure generalizes the case of the
 massive superparticle as obtained from the massless
 superparticle in one higher dimension, by viewing the mass as an
 integration constant of this last one E-L equations.
 Specifically,

 \noindent 1) We present a `tension generating mechanism'
 in a kind of first order formulation that extends the
 true first order formulation of the $p=0$ case
 ({\it cf.} \cite{To92,BLT92}).

\noindent 2) We do not include in the super-$p$-brane action any
 higher form fields directly defined on the worldvolume, only the auxiliary
 scalar ones. We achieve this by suitably enlarging ordinary superspace
 {\it i.e.}, by adhering to the enlarged superspace
 variables/worldvolume fields correspondence for branes
 \cite{ACIP-00} as it is also done for the D-branes in
 \cite{ACIP-00,Saka}.

\noindent 3) The use of the above enlarged superspaces allows
 us to characterize $\kappa$-symmetry transformations
 as local, $\kappa$-dependent right translations associated
 with the corresponding extended superspace group coordinates, as it is the case
 for the standard superspaces. $\kappa$-symmetry gauge transformations
 also include in our case $\kappa$-dependent bosonic transformations.

\noindent 4) The new super-$p$-brane actions on the enlarged
superspaces possess strict $\kappa$-invariance: their Lagrangian
is itself $\kappa$-invariant. This requires that the additional
field variables in $\tilde\Sigma$ be non-inert under
$\kappa$-transformations. This strict $\kappa$-invariance is
analogous to the strict supersymmetry invariance
\cite{BeSe95,ACIP-00} of the standard Lagrangians (WZ terms)
constructed on the enlarged superspaces. Strict
$\kappa$-invariance is, however, optional for these
super-$p$-brane actions (which explains why the new variables in
$\tilde\Sigma$ may be inert under $\kappa$-symmetry, as in
\cite{BeSe95}) and is a necessary result for the new actions and
Lagrangians discussed here.

Finally, we mention that our results may be generalized to $p>2$.
The actions (\ref{B10}) and (\ref{Me8a}) can be extended to higher
$p$ (a) using the appropriate higher $p$ enlarged superspaces for
the contribution of $\Phi$ from a manifestly invariant WZ form and
(b) keeping the same structure of (\ref{B10}) and (\ref{Me8a}) but
now with auxiliary variables $p^{\mu_1\dots\mu_{p+1}}$,
$E=\frac{1}{(p+1)!}e_{i_1\dots i_{p+1}}d\xi^{i_1}\dots
d\xi^{i_{p+1}}$.

\bigskip

{\it Acknowledgments}. This work has been partially supported by
the research grants BFM2002-03681 and BFM2002-02000 from the
Spanish Ministerio de Educaci\'on y Ciencia, VA085-02 from the
Junta de Castilla y Le\'on and from EU FEDER funds. C.M.-E. wishes
to thank the Ministerio de Educaci\'on y Ciencia for his FPU
research grant. Discussions with I. Bandos and D. Sorokin are
gratefully acknowledged.

\end{document}